# Encompression Using Two-dimensional Cellular Automata Rules

**[1]Sudhakar Sahoo, [2]Sanjaya Sahoo, [3]Birendra Kumar Nayak and [4]Pabitra Pal Choudhury**

[1, 4]Applied Statistics Unit, Indian Statistical Institute, Kolkata, 700108, INDIA
Email: sudhakar.sahoo@gmail.com, pabitrapalchoudhury@gmail.com
[1]Gandhi Institute for Technical Advancement, Jaanla, Bhubaneswar, INDIA
Email: sanjayamtech@gmail.com
[3]P.G.Department of Mathematics, Utkal University, Bhubaneswar-751004
Email: bknatuu@yahoo.co.uk

*Abstract-* **In this paper, we analyze the algebraic structure of some null boundary as well as some periodic boundary 2-D Cellular Automata (CA) rules by introducing a new matrix multiplication operation using only AND, OR instead of most commonly used AND, EX-OR. This class includes any CA whose rule, when written as an algebra, is a finite Abelean cyclic group in case of periodic boundary and a finite commutative cyclic monoid in case of null boundary CA respectively. The concept of 1-D Multiple Attractor Cellular Automata (MACA) is extended to 2-D. Using the family of 2-D MACA and the finite Abelian cyclic group, an efficient encompression algorithm is proposed for binary images.**

## I. INTRODUCTION

Real time secured compression is all the time needed for various confidential transactions. Cryptography, to most people, is concerned with keeping communications private. The effective measure of cryptosystem is how long it can be used to encrypt and decrypt images without the key being broken. On the other hand, lossy data compression is a process of reducing the amount of data required to represent a given quantity of information with acceptable loss. It removes redundancy, repeatability and irrelevancy of data blocks of input file to generate the compressed output. Encompression is a process of combining both encryption as well as compression in a single platform. D-encompresion is a reverse process of encompresion. Cellular Automata (CA) based encompression algorithm presents one promising approach to both compression and encryption.

In the past, Lafe [2] used Cellular Automata Transforms (CAT) for encryption and compression of both text as well as the image data. Shaw et al. [6] uses 1-D Multiple Attractor Cellular Automata (MACA) and proposed a CA based encompression technology for voice data.

In this paper, a different encompression model is proposed for binary images. Here a binary image is considered as the initial state of a 2-D CA and the triplets (dimension of the image, a 2-D MACA rule, an invertible rule matrix) is used as a key for the encompression process, evolving an unpredictable complex chaotic system from this 'initial state'. Here we have used symmetric key cryptosystem, which means that, the sender and the receiver both share a common key through a secure channel. The sender uses the key to encrypt the compressed image, which he wants to send. In the D-encompresion part the receiver decrypts and decompresses the image using the same key (used by sender) to get back a lossy image. The major contribution of this paper can be summarized as follows.

. A **finite Abelian cyclic group** is obtained for 2-D nine neighborhood periodic boundary CA rules, by introducing a new multiplication operation on rule matrices using only AND, OR.

. For null boundary CA, the same operation gives a **finite commutative cyclic monoid.**

. The concept of 1-D MACA and their properties are extended to 2-D.

. Both 2-D MACA and the CA rules obtain from the finite Abelian group are used for Encompression (Encryption and Compression) of a binary image.

Moore [8, 9] pointed out that if the CA has certain algebraic properties then the computational complexity of an algorithm reduces. So, apart from the encompression application, the algebraic structures [a finite Abelian cyclic group and a finite commutative cyclic monoid] obtained from our research can also be used to solve the CA prediction problem more quickly even for some non-linear rules.

Section II gives the basic concept of Cellular Automata. The issues regarding the extension from 1-D MACA to 2-D are discussed in section III. The algebraic structures obtained from the set of basic rule matrices by introducing a new matrix multiplication operation is shown in Section IV. Using this family of CA rules and 2-D MACA an encompression scheme is presented in section V. Finally a conclusion is drawn in Section VI.

## II. BASIC CONCEPTS

Throughout this paper, *Logical OR* operation is denoted by $'\vee'$, *AND* by $'\wedge'$, *EX-OR* by $'\oplus'$, arithmetic addition by $'+'$ and arithmetic multiplication by $'\times'$.

A Cellular Automata (CA) [1] is an idealized parallel processing machine consisting of a number of cells containing some cell values called states together with an updating rule. A cell value is updated based on this updating rule, which involves the cell value as well as other cell values in a particular neighborhood. In this paper we have concentrated only on 1-D and 2-D elementary CA having the cell values either 0/ 1.

The *global state* or simply *state* of a 1-D binary CA at any time-instant $t$ is represented as a vector $X^t = (x_1^t, x_2^t, \ldots, x_n^t)$ where $x_i^t$ denotes the bit in the $i^{th}$ cell $x_i$ at time-instant $t$. However, instead of expressing a state as a bit-string, we shall frequently represent it by the decimal equivalent of the $n$-bit string with $x_1$ as the Most Significant Bit; e.g. for a 4-bit CA, the state *1011* may be referred to as state 11 ($=1\times2^0 +1\times2^1 +0\times2^2 +1\times2^3$).

The bit in the $i^{th}$ cell at the "next" time-instant $t+1$ is given by a *local mapping* denoted by $f^i$, say, which takes as its argument a set of the bits at time-instant t in the cells of a certain pre-defined *neighborhood* (of size $p$, say) of the $i^{th}$ cell. In *elementary 1-D CA* defined by Wolfram [3], a symmetrical neighborhood of size 3 is used for each cell so that $x_i^{t+1} = f^i(x_{i-1}^t, x_i^t, x_{i+1}^t)$ , $i = 2,3,\ldots\ldots, n-1$.

DEFINITION 1: A null boundary CA is the one in which the extreme cells are connected to logic - 0 states.

DEFINITION 2: A periodic boundary CA is the one in which the extreme cells are connected to each other.

A 1-D CA may be represented as a string of the rules applied to the cells in proper order, along with a specification of the boundary conditions. e.g. <103, 234, 90, 0>NB refers to the CA $(x_1, x_2, x_3, x_4)$ where $x_1^{t+1} = f_{103}(0, x_1^t, x_2^t)$; $x_2^{t+1} = f_{234}(x_1^t, x_2^t, x_3^t)$; $x_3^{t+1} = f_{90}(x_2^t, x_3^t, x_4^t)$; $x_4^{t+1} = f_0(x_3^t, x_4^t, 0)$.

If the "present state" of an $n$-bit CA (at time $t$) is $X^t$, its "next state" (at time $t+1$), denoted by $X^{t+1}$, is in general given by the *global mapping* $F(X^t)=(f^1(lb^t, x_1^t, x_2^t), f^2(x_1^t, x_2^t, x_3^t), \ldots, f^n(x_{n-1}^t, x_n^t, rb^t))$, where $lb$ and $rb$ denote respectively the left boundary of $x_1$ and right boundary of $x_n$.

If the rule applied to each cell of a CA is a linear Boolean function, the CA will be called a Linear Cellular Automaton, otherwise a Non-linear Cellular Automaton, e.g.<0, 60, 60, 204>NB is a linear CA while <60,90,87,123>PB is a non-linear CA.

If the same Boolean function (rule) determines the "next" bit in each cell of a CA, the CA will be called a Uniform Cellular Automaton (UCA), otherwise it will be called a Hybrid Cellular Automaton (HCA), e.g.<135, 135, 135, 135>PB is a UCA, <0, 60, 72, 72>NB is a HCA.

For a UCA, the Boolean function applied to each cell will be called the rule of the CA. So for a UCA, we can simply denote it as $f$. e.g. for the 4-bit CA <230, 230, 230, 230>PB, the rule of the CA is Rule 230 and the CA will be called the "Rule 230 CA" of 4 bits with periodic boundary conditions. In 2-D Nine Neighborhood elementary CA: the next state of a particular cell is obtained by the current state of itself and eight cells in its nearest neighborhood (also referred as Moore neighborhood) as shown in Fig-1. Such dependencies are accounted by various rules or transition functions. For the sake of simplicity, in this section we take into consideration only the linear rules, i.e. the rules, which can be realized by $'\oplus'$ operation only. A specific rule convention that is adopted in our previous paper [5] is as follows:

| 64 | 128 | 256 |
|----|-----|-----|
| 32 | 1 | 2 |
| 16 | 8 | 4 |

[Fig-1]

The central box represents the current cell (i.e. the cell being considered) and all other boxes represent the eight nearest neighbors of that cell. The number within each box represents the rule number characterizing the dependency of the current cell on that particular neighbor only. Rule 1 characterizes dependency of the central cell on itself alone whereas such dependency only on its top neighbor is characterized by rule 128, and so on. These nine rules are called fundamental rules. In case the cell has dependency on two or more neighboring cells, the rule number will be the arithmetic sum of the numbers of the relevant cells. For example, the 2D CA rule 171 (128+32+8+2+1) refers to the five-neighborhood dependency of the central cell on (top, left, bottom, right and itself). The number of such rules is 512, which also includes the rule characterizing no dependency.

Both 1-D and 2-D linear CA rules can be characterized using a transformation matrix denoted by $T_R$ such that $T_R$ when multiplied on the current CA states X (n-bit CA incase of 1-D and a (m x n) binary matrix incase of 2-D) generates the next state X′. The dimension of $T_R$ is (n x n) incase of 1-D and it is (mn x mn) in case of 2-D [5].

A special class of CA termed as Multiple Attractor Cellular Automata (MACA) consists of a number of cyclic and non-cyclic states when drawing its State Transition Diagram (STD). The STD for a1-D, 5-cell, MACA for the rule vector <102, 60, 204, 204, 170> is shown in fig 2.

DEFINITION 3: An m-bit field of an n-bit attractor is said to be pseudo-exhaustive if all possible $2^m$ states appear in the set.

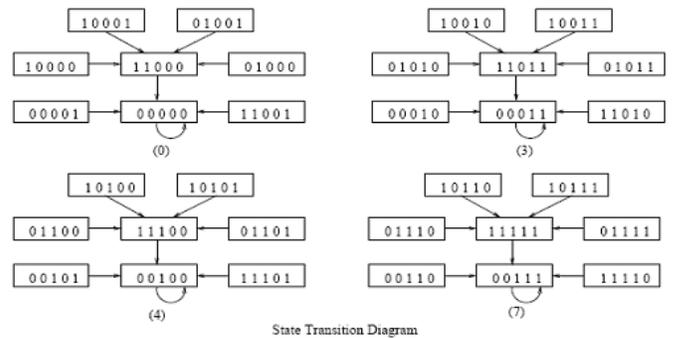

State Transition Diagram

[Fig 2: Shows an STD of a 1-D MACA]

**Theorem 1:** [4] In an n-cell MACA with k=2^m attractors, there exists m-bit positions at which the attractors generate pseudo-exhaustive 2^m states. Further, the number of predecessors in a MACA is 2^{n-r}, where n is the number of cells of the CA and r is the rank of its transformation matrix $T_R$.

The pseudo exhaustive field (PEF) of an attractor provides the pointer to the class of states in the attractor basin. In order to identify the class of a state *p*, the MACA is initialized with *p* and operated maximum of d (depth) number of cycles till it reaches an attractor. Next, the PEF bits can be extracted (as noted in [4, 7] ) to identify the class of *p*. The detail characterization of 1-D MACA is available in [4, 7].

The MACA having depth equal to one is called D1-MACA. For any d-depth MACA there exists a D1-MACA [10]. In a single iteration D1-MACA can reach at the attracter and drastically reduces the time complexity.

## III. MACA IN TWO DIMENSION

Just like 1-D MACA, 2-D MACA can also be analyzed in a similar way. Here one may get different clusters (called Attractor Basins) containing attractors. Starting from a binary image of size (m x n), the number of steps required to reach at the attractor is called the DEPTH. We have observed that, most of the characteristics of 1-D MACA and 2-D MACA are consistent. Similar to 1-D the, the PEF bits of attractors in 2-D can also be used to save the memory and is the basis for the compression part used in this paper. Both the algorithms O(n^3) [4] and O(n) [7] can be used to identify the PEF of an attractor basin. Below we give an example of a 2-D MACA in two-dimension by applying a CA rule to a (2 x 2) matrix. In 2-D MACA, the numbering of states are the decimal values computed in a row major order or column major order, diagonals etc.

**Example 1:** Fig 3 shows an example of a 2-D MACA for the linear rule 69 ((1+4+64) see fig-1 for their positions) in Null boundary condition. Here we are getting 4 attractors as 0, 2, 4 and 6. The row major order convention is used to find out the state number. State 0 in decimal means that, a (2 x 2) matrix $\begin{pmatrix} 0 & 0 \\ 0 & 0 \end{pmatrix}$, the state $2 = \begin{pmatrix} 0 & 0 \\ 1 & 0 \end{pmatrix}$, for state 4 the matrix is $\begin{pmatrix} 0 & 1 \\ 0 & 0 \end{pmatrix}$ and for the state 6 the matrix is $\begin{pmatrix} 0 & 1 \\ 1 & 0 \end{pmatrix}$. Other state numbers can also be given in a similar way. For the attractors, the PEF bits are present in the diagonal cells. Except these 4 attractors, all other states are non-cyclic states. The states 1, 3, 5, 8, 10, 12 and 14 are non-reachable states and their depth is 2. The depth of 9, 11, 13 and 15 is equal to 1. The transformation matrix (or Rule matrix in 2-D) for this linear rule is given below.

$$M_{69} = \begin{pmatrix} 1 & 0 & 0 & 1 \\ 0 & 1 & 0 & 0 \\ 0 & 0 & 1 & 0 \\ 1 & 0 & 0 & 1 \end{pmatrix}.$$

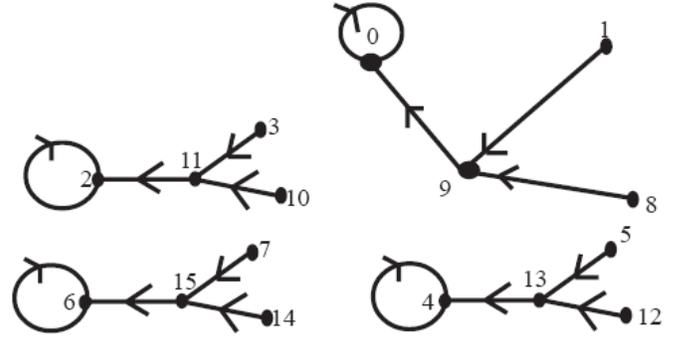

[Fig 3: Shows an STD of a 2D MACA rule 68]

## IV. ALGEBRAIC STRUCTURE OF CA RULES

A quick review of our previous work for the easy construction of rule matrices for null boundary and periodic boundary CA follows.

### RULE MATRIX IN NULL BOUNDARY CA

If the problem domain is a matrix of order (m x n) then the rule matrices are of order (mn x mn). Let $S_1$ and $S_2$ be two sequences of '0' and '1' defined as follows:

$$S_1 = \underbrace{1111...1}_{(n-1)} 0 \underbrace{1111...1}_{(n-1)} 0... \quad ...0 \underbrace{1111...1}_{(n-1)} 0 \underbrace{111...1}_{(n-1)}$$

i.e. the sequence contains (n-1) '1'followed by a'0', then (n-1) '1' and '0' and finally end with(n-1)'1' and

$$S_2 = 0 \underbrace{111...1}_{(n-1)} 0 \underbrace{111...1}_{(n-1)}... \quad ...0 \underbrace{111...1}_{(n-1)} 0 \underbrace{111...1}_{(n-1)}$$

i.e. the sequences starts with a '0' followed by a string of (n-1)'1'then a '0' and again followed by (n-1)'1' and so on. Where n= number of columns in the given problem matrix. The 5 basic matrices possess a particular structure as follows.
$M_1$=main diagonal contains all '1's and other elements are '0'.
$M_2$=super diagonal elements are arranged in a sequence like $S_1$ and all other elements are '0'
$M_4$= $(n+1)^{th}$ diagonal elements are arranged in a sequence like $S_1$ and all other elements are '0'
$M_8$=n^{th} diagonal contains all'1's and all others are'0'.
$M_{16}$= $(n-1)^{th}$ diagonal elements are arranged in a sequence like $S_2$ and all other elements are'0'.

Just like the rule matrices for null boundary CA, we have observed similar results for the periodic boundary CA as follows.

### RULE MATRIX IN PERIODIC BOUNDARY CA

Let us define two matrices $T_1$ and $T_2$ as follows:

$$T_1 = \begin{pmatrix} 0 & 0 & 0 & 1 \\ 1 & 0 & 0 & 0 \\ 0 & 1 & 0 & 0 \\ 0 & 0 & 1 & 0 \end{pmatrix} \text{ and } T_2 = (T_1)^T = \begin{pmatrix} 0 & 1 & 0 & 0 \\ 0 & 0 & 1 & 0 \\ 0 & 0 & 0 & 1 \\ 1 & 0 & 0 & 0 \end{pmatrix}$$

The rule matrices of order (mn x mn) can be partitioned into (n x n) smaller matrices arranged in m rows and m columns as shown below.

$$M_R = \begin{pmatrix} M_1^1 & M_2^1 & . & . & M_m^1 \\ M_1^2 & M_2^2 & . & . & M_m^2 \\ . & . & . & . & . \\ . & . & . & . & . \\ M_1^m & M_2^m & . & . & M_m^m \end{pmatrix}$$

Where $M_k^i$ is an (n x n) matrix, $1 \le i, k \le m$. The structure of rule matrices for 5 basic rules in periodic boundary CA is as follows.

$M_1$: $M_i^i$ are identity matrices for i=1, 2...m.

$M_2$: $M_i^i$ are of type $T_2$, for i=1, 2...m.

$M_4$: Super diagonals and (m-1)$^{th}$ sub diagonal are of type $T_2$

$M_8$: Super diagonals and (m-1)$^{th}$ sub diagonal are the identity matrices of order(n x n)of the above partitioned matrix.

$M_{16}$: Super diagonals are of type $T_1$ and (m-1)$^{th}$ sub diagonal are of type$T_2$

**Theorem 2**: The fundamental matrices $M_1$, $M_2$, $M_4$, $M_8$, $M_{16}$, $M_{32}$, $M_{64}$, $M_{128}$ and $M_{256}$ in 2-D periodic boundary CA are all non-singular.

**Proof:** $M_1$ is the identity matrix. Hence it is non-singular.
Again by inter changing different rows of the fundamental matrices one can reduce it to identity matrix (or rule 1 matrix). Hence they are also non-singular. Hence proved.

ALGEBRIC STRUCTURE OF NULL BOUNDARY CA

Let M= {$M_1$, $M_2$, $M_4$, $M_8$, $M_{16}$} be the set of 5 basic matrices and ' *' be the multiplication operation defined on M as follows.
If A=($a_{ij}$) and B=($b_{ij}$) are two matrices of M then A * B = C

Where C=$c_{ij}$ and $c_{ij} = \bigvee_{k=1}^{mn} (a_{ik} \wedge b_{kj})$ for all $1 \le i, j \le mn$.

Let $M'$ = {M, *} be a class generated on multiplying some or all of the matrices of M. If M contains the basic NULL BOUNDARY CA RULES, then it can be easily verified that the set $M'$ will satisfy some algebraic properties such as CLOSURE, ASSOCIATIVITY, COMMUTATIVE and the EXISTANCE OF IDENTITY. Thus

**Theorem 3**: The set $M'$ = (M, *) is a FINITE COMMUTATIVE CYCLIC MONOID and the 5 basic rule matrices are the generators.

ALGEBRIC STRUCTURE OF PERIODIC BOUNDARY CA RULES

Let $M''$ = (M, *) be a class generated on multiplying some or all of the matrices of M. If M contains the basic PERIODIC BOUNDARY CA RULES then it can be easily verified that $M''$ will satisfy CLOSURE, ASSOCIATIVITY, INVERSE, COMMUTATIVE and the EXISTANCE OF IDENTITY. Therefore

**Theorem 4**: The set $M''$ = (M, *) is a FINITE ABELIAN CYCLIC GROUP where the 5 basic rule matrices are the generators.

**Theorem 5**: If the problem matrix is of order (m x n) then $o(M') = o(M'') = mn$. Where $o(S)$ is defined as the order or the number of CA rules present in the set S, also known as the cardinality of $S$.

**Proof:** As the 5 basic rule matrices are all lower triangular matrices, so the multiplication of any matrices between them shifts the 1's in diagonal wise. This characteristic of matrix multiplication is reported in our previous paper [5] as a translation of images in different directions. As maximum number of diagonal in a rule matrix is of order (mn x mn), thus the number of different matrices possible is mn. Hence proved.

Another important observation is that, if mn > 9 then all the 9 fundamental linear rules $M_1$, $M_2$, $M_4$, $M8$, $M_{16}$, $M_{32}$, $M_{64}$, $M_{128}$ and $M_{256}$ are present in $M''$. Except these 9 rules other rules in $M''$ are all linear in $p$ neighborhood 1-D CA with $p$>3.

V. ENCOMPRESSION AND D-ENCOMPRESSION

In this section we introduce the technique of encompression. Encompression means both encryption as well as compression. When one wants to send an image; first it is compressed then encrypted, so that when the receiver receives the image he/she gets an encompressed image. In section IV, the finite Abelian cyclic group containing invertible matrices is obtained which is the basis for both encryption and decryption. But at the time of compression and decompression, the 2D-MACA rules discussed in section III is used. Here we use symmetric encryption technique. Both the sender and receiver know the triplet: (**dimension of the image, a 2-D MACA rule, an invertible rule matrix from the finite Abelian cyclic group**) which is used as a key at the time of both compression and encryption of an image. The proposed encompression and D-encompresion model is shown in fig 4.

• ENCOMPRESSION OF A BINARY IMAGE

COMPRESSION PART

Starting from a binary image of order (m x n) a 2-D MACA having k attractors can run for d (depth) times (d=1 for D1 MACA) to reach at the attractor, which is necessarily a binary matrix of order (m x n). For compression we shall find out the PEF bits from the attractor matrices of order (m x n) using the O(n) algorithm given in [7]. This requires $\lceil \lg k \rceil$ bits and this is the size of the compressed image. The compression ratio in this scheme is $\lceil \lg k \rceil / mn$.

ENCRYPTION PART

The PEF bits cab be arranged to get a binary matrix of order (p x q). Then we multiply a reversible CA rule matrix from the class of finite Abelian cyclic group and can get a matrix of order (p x q). That is our encompressed image.

• D-ENCOMPRESSION OF A BINARY IMAGE

## DECRYPTION PART

When the receiver receives the encompressed image of order (p x q) first he/she decrypts the image by using the inverse of the reversible CA rule and gets back an image of order (p x q). This is the PEF bits of the attractors of 2-D MACA used in the encompression process. Then he/she goes for decompression.

## DECOMPRESSION PART

After getting the attractor the receiver can compute all its predecessors preset in an attractor basin. Thus the receiver can able to know some matrices, which belong to a cluster and consider arbitrarily one matrix from it. This matrix may or may not be the original image matrix but matrices of one cluster are very similar in the sense that the hamming distance between the matrices is very less. For that reason we shall get a lossy image with minimum error.

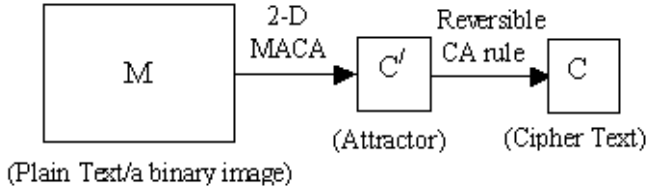

(a) Shows an encompression scheme for a binary image

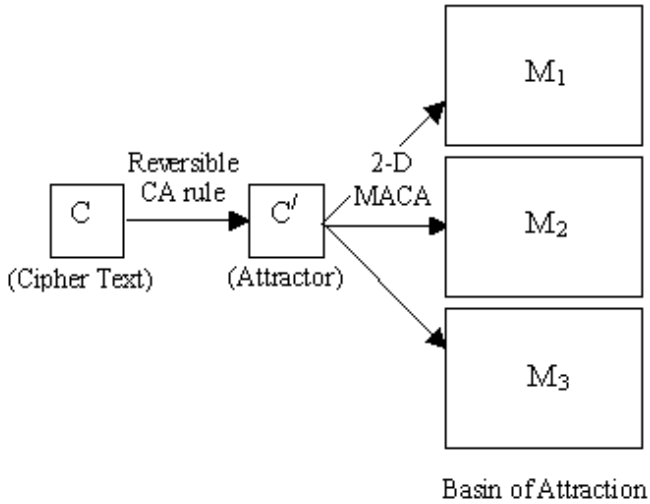

(b) The D-encompresion process which is basically a lossy procedure with minimum error.

[Fig 4: Proposed Encompression model for binary images.]

## STRENGTH OF THE ENCRYPTION PART

1. In a (m x n) matrix total number of entries is mn. So for compression, attacker can approach in $(512)^{mn}$ ways to get the rule. Which is exponential.
2. For decryption he/she requires mn number of ways to get the rule.

3. So in a brute force approach, $\{(512)^{mn}$ x mn $\}$trials are required for an attacker to get the rules for decryption and decompression of the image.

## VI. CONCLUSION

In this paper, we analyze the null boundary and periodic boundary CA rules from 5 basic rule matrices and introduces a new matrix multiplication operation using only AND, OR instead of using AND, XOR. The non-linear operator OR which when replaces the linear operator XOR and acts on the domain of linear rules interesting algebraic structures are obtained in different boundary conditions. Out of this one is a finite Abelean cyclic group and the other one is a finite commutative cyclic monoid. Further, we deeply analyze the concept of 2-D MACA as an extension over 1-D MACA. We have also used the 2-D CA rules taken from the finite Abelian cyclic group and from the 2-D MACA for both compression as well as encryption of a binary image. Parallelism behavior inherently presents in the concept of CA and the creation of attractor in 2-D helps us to reduce the time complexity as well as the memory overhead for the proposed encompression technique.

Also the theoretical results obtained during our research in finding a class of reversible CA rules in 2-D must help other researchers to work in it.